\documentclass[conference]{IEEEtran}
\usepackage{cite}
\usepackage{amsmath,amssymb,amsfonts}
\usepackage{algorithmic}
\usepackage{graphicx}
\usepackage{textcomp}
\usepackage{pifont}
\usepackage[dvipsnames]{xcolor}
\usepackage[para]{footmisc}
\usepackage{listings}
\usepackage{xcolor}

\definecolor{codegreen}{rgb}{0,0.6,0}
\definecolor{codegray}{rgb}{0.5,0.5,0.5}
\definecolor{codepurple}{rgb}{0.58,0,0.82}
\definecolor{backcolour}{rgb}{0.95,0.95,0.92}

\usepackage[english,linesnumbered,vlined,ruled,nokwfunc]{algorithm2e}
\DontPrintSemicolon
\SetKwComment{note}{$\triangleright$ }{}
\SetFuncSty{textsc}
\SetCommentSty{textit}
\SetDataSty{texttt}
\SetKwInput{Initial}{Initial}
\SetKwInput{Parameter}{Param.}
\SetKwInput{Input}{Input}
\SetKwInput{Data}{Data}
\SetKwInput{Output}{Output}
\ResetInOut{Result} 
\let\oldnl\nl%
\newcommand{\nonl}{\renewcommand{\nl}{\let\nl\oldnl}}%
\SetKw{KwAnd}{and} %
\SetKw{KwOr}{or}
\SetKw{KwXor}{xor}
\SetKw{KwNot}{not}
\SetKw{Parallel}{parallel}
\SetKw{Return}{return}

\usepackage{graphicx}
\newcommand{\tg}{\mathcal{G}}
\newcommand{\tge}{\mathcal{E}}
\newcommand{\alltimes}{{T}}

\newcommand{\tglib}{\textsc{TGlib}\xspace}
\newcommand{\dseil}{\textsc{Ilists}\xspace}
\newcommand{\dsstr}{\textsc{Stream}\xspace}
\newcommand{\dstra}{\textsc{Trs}\xspace}
\newcommand{\dsdlg}{\textsc{Dlg}\xspace}
\newcommand{\dsagg}{\textsc{Aggr}\xspace}
\usepackage{xspace}

\usepackage{booktabs}
\usepackage{multirow}
\usepackage{siunitx}
\usepackage{graphicx}
\usepackage{multirow}
\usepackage{skull}
\usepackage{amsmath}
\usepackage{amssymb}
\usepackage{amsfonts}
\usepackage{amsthm}
\usepackage{thmtools}		
\usepackage{mleftright}
\usepackage{thm-restate}
\usepackage[mathic=true]{mathtools}
\usepackage{fixmath} 
\usepackage{paralist}
\lstdefinestyle{mystyle}{
    backgroundcolor=\color{backcolour},   
    commentstyle=\color{codegreen},
    keywordstyle=\color{magenta},
    numberstyle=\tiny\color{codegray},
    stringstyle=\color{codepurple},
    basicstyle=\ttfamily\footnotesize,
    breakatwhitespace=false,         
    breaklines=true,  
    postbreak=\mbox{\hspace{2mm}$\hookrightarrow$\space},               
    captionpos=b,                    
    keepspaces=true,                 
    numbers=left,                    
    numbersep=5pt,                  
    showspaces=false,                
    showstringspaces=false,
    showtabs=false,                  
    tabsize=2
}

\newtheorem{definition}{Definition}
\usepackage{url}
\usepackage[hidelinks]{hyperref}
\usepackage{cleveref}
\theoremstyle{remark}

\usepackage{tikz}
\usetikzlibrary{arrows,decorations.pathmorphing}
\usetikzlibrary{arrows.meta}
\tikzset{
    vertex/.style={circle,draw,inner sep=01pt,minimum size=1.3em},
    vertex2/.style={circle,draw,inner sep=01pt,minimum size=1.5em},	
    vertexa/.style={circle,draw,fill=red, minimum size=2mm,inner sep=1pt, font=\scriptsize},	
    vertexb/.style={circle,draw,fill=black, minimum size=2mm,inner sep=1pt, font=\scriptsize},	
    vertexb2/.style={draw,inner sep=01pt,minimum size=1.3em},
    vertexc/.style={circle,draw,fill=green, minimum size=2mm,inner sep=1pt, font=\scriptsize},	
    vertexd/.style={circle,draw,fill=black, minimum size=2mm,inner sep=0pt},
    vertexldg/.style={circle,draw,fill=gray, minimum size=2mm,inner sep=0pt,color=gray},		
    vertexld/.style={draw,fill=white, minimum size=4.1mm,inner sep=0pt},	
    ldgedge/.style={->,> = latex', font=\footnotesize,color=gray,dashed},
    dedge/.style={->,> = latex', font=\footnotesize},
    wedge/.style={->,> = latex', font=\footnotesize, dashed},
    edge/.style={-,> = latex', font=\footnotesize},
}

\begin{document}

\title{TGLib: An Open-Source Library for Temporal Graph Analysis}

\author{\IEEEauthorblockN{Lutz Oettershagen}
    \IEEEauthorblockA{\textit{University of Bonn}}
    Bonn, Germany\\
    \url{lutz.oettershagen@cs.uni-bonn.de}\\
    \and
    \IEEEauthorblockN{Petra Mutzel}
    \IEEEauthorblockA{\textit{University of Bonn}}
    Bonn, Germany\\
    \url{petra.mutzel@cs.uni-bonn.de}\\
    
    }
\maketitle

\begin{abstract}\boldmath
We initiate an open-source library for the efficient analysis of temporal graphs. We consider one of the standard models of dynamic networks in which each edge has a discrete timestamp and transition time. Recently there has been a massive interest in analyzing such temporal graphs. Common computational data mining and analysis tasks include the computation of temporal distances, centrality measures, and network statistics like topological overlap, burstiness, or temporal diameter. To fulfill the increasing demand for efficient and easy-to-use implementations of temporal graph algorithms, we introduce the open-source library \tglib, which integrates efficient data structures and algorithms for temporal graph analysis. \tglib is highly efficient and versatile, providing simple and convenient C++ and Python interfaces, targeting computer scientists, practitioners, students, and the (temporal) network research community.
\end{abstract} 

\section{Introduction}

\tglib is an open-source C++ template library with an easy-to-use Python front-end focusing on efficient temporal graph analysis tasks.
Network data often originates from dynamic systems that change over time: Links are formed or broken, such that the topology of the network changes over time. 
\emph{Temporal graphs} capture these changes. A temporal graph is a graph that changes over time, i.e., each edge has a time stamp that determines when the edge exists in the graph. Hence, the topology of the graph changes in discrete time steps.
Temporal graphs are often good models for real-life scenarios due to the inherently dynamic nature of most real-world activities and processes. In many situations, events, e.g., communication in social networks, are time-stamped, such that temporal graphs naturally arise from the recorded data. 

A finite temporal graph consists of a finite set of (static) vertices and a finite set of temporal edges. 
A temporal edge connects two vertices at a discrete \emph{availability time}, and edge traversal costs a (usually) non-negative amount of time (called the \textsl{transition time}). The availability time denotes the time when an edge is available for transition, and the transition time defines how long the transition takes. 

This work introduces \tglib, an open-source toolkit for handling and analyzing temporal graphs. Our library is geared towards a variety of research communities that often need to work with increasingly larger temporal graphs. 
The applications of temporal graphs are manifold:
\newline
\emph{--Communication networks} are a prime example of the application of temporal graphs.
Email (or text message) networks model the (almost) instantaneous communication between the participants, and have been used to identify 
different dynamics of communication as well as properties of the participants~\cite{candia2008uncovering,eckmann2004entropy,holme2004structure}. 
Vertices of the network represent the participants and temporal edges represent each communication. 
\newline
\emph{--Proximity and contact networks} record the contacts between individuals
by measuring their proximity. For example, modern
smartphones are ubiquitous and can record the proximity of users to identify contacts or build opportunistic networks~\cite{avin2008explore,Chaintreau2007}.
Several works discuss the spreading of diseases in contact networks, e.g.,~\cite{ciaperoni2020relevance,oettershagen2020classifying}.
\newline
\emph{--Social networks}, formal or informal, are a fundamental part of human life.
Nowadays, online social networks, like \emph{Facebook} or \emph{WeChat}, host billions
of users. In online and offline social networks, participants join and leave the network over time and form or end relations with each other. Recent works discuss the importance of temporal properties, e.g.,~\cite{hanneke2006discrete,moinet2015burstiness}.
\begin{figure}
    \centering
    \includegraphics[width=0.9\linewidth]{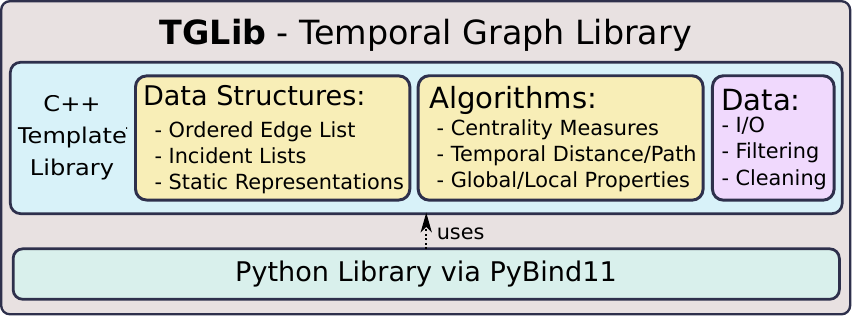}
    \caption{Overview of the high-level architecture of the \tglib library. The data structures and temporal graph algorithms are implemented in efficient C++. A Python front-end provides easy-to-use access.}
    \label{fig:overview}
\end{figure}

There are many more prominent use-cases for temporal graphs, like modeling transportation networks~\cite{gallotti2015multilayer,Huynh2021} or applications in biology, e.g., modeling dynamic protein-protein interactions~\cite{lebre2010statistical,przytycka2010toward}, and neural brain networks~\cite{thompson2017static}.

\noindent\textbf{Contributions:~}
We introduce \tglib, an open-source temporal graph library under the permissive MIT license.
Our library focuses on temporal distance and centrality computations and other local and global temporal graph statistics.
\tglib is designed for performance and usability by an efficient and modular C++ implementation of the core data structures and algorithms and an easy-to-use Python front-end allowing users and researchers without in-depth (C++) programming experience to use our new library.

Additionally, we offer the first implementations of new algorithms for the earliest arrival distance using the temporal graph data structure introduced in~\cite{gheibi2021effective}, a variant of the fastest path algorithm from~\cite{oettershagen2020efficient} for shortest temporal paths,
a \mbox{top-$k$} harmonic closeness algorithms for shortest temporal path distance based on the algorithmic ideas of the top-$k$ closeness algorithm using minimum duration distance~\cite{oettershagen2020efficient}, and finally, an algorithm for the temporal edge betweenness based on the directed line graph representation~\cite{oettershagen2020temporal}.

\section{Preliminaries}

A \emph{temporal graph} $\tg=(V,\tge)$ consists of a finite set of vertices $V$ and a finite set of directed \emph{temporal edges} $\tge$.
A temporal edge $e=(u,v,t,\lambda)$ consists of the vertices $u,v\in V$, \emph{availability time} (or \emph{time stamp}) $t \in \mathbb{N}$ and \emph{transition time} $\lambda \in \mathbb{N}$, i.e., $e=(u,v,t,\lambda)\in V\times V \times \mathbb{N}^2$.
We model an undirected temporal graph by a directed temporal graph
using a forward- and a backward-directed edge with equal time stamps and traversal times for each undirected edge.

We use $n=|V|$ and $m=|\tge|$ to denote the numbers of vertices and temporal edges, respectively.
The \emph{arrival time} of an edge $e=(u,v,t,\lambda)$ (at vertex $v$) is $t+\lambda$.
We use $V(\tg)$ to denote the set of vertices of $\tg$,
the minimal number of incoming (outgoing) temporal edges over all vertices by $\delta^-_{min}$ ($\delta^+_{min}$). Similarly, we define $\delta^-_{max}$ ($\delta^+_{max}$) for the maximal incoming (outgoing) degree.
Furthermore, we denote with $N(u)=\{v\mid (u,v,t,\lambda)\in\tge\}$ the neighborhood of $u$.
%
%
With  $\alltimes(\tg)$, we denote the set of all availability times of edges in $\tg$, i.e., $\alltimes(\tg)=\{t\mid (u,v,t,\lambda)\in \tge \}$.
Given a temporal graph $\tg=(V,\tge)$, it is common to restrict algorithms and computations on $\tg$ to a given \emph{restrictive time interval}
$I=[a,b]$ with $a,b\in\mathbb{N}$, such that only the temporal subgraph $\tg'=(V,\tge')$ with $\tge'=\{(u,v,t,\lambda)\in \tge \mid t\geq a \text{ and } t+\lambda \leq b\}$ needs to be considered.

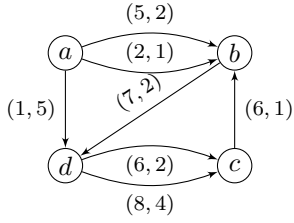
\begin{figure}[t]
    \centering
    \begin{tikzpicture}[scale=0.75]
        \node[vertex] (a) at (0,0) {$a$};
        \node[vertex] (b) at (3,0) {$b$};
        \node[vertex] (c) at (3,-2){$c$};
        \node[vertex] (d) at (0,-2) {$d$};
        \path[->] 	
        (a)  edge[bend right=20, dedge]  node[above] {$(2,1)$}   (b)
        (a)  edge[bend left=20, dedge]  node[above] {$(5,2)$}   (b)
        (a)  edge[dedge]  node[midway,left] {$(1,5)$}   (d)
        (b)  edge[dedge]  node[sloped,midway,above] {$(7,2)$}  (d)
        (c)  edge[dedge]  node[midway,right] {$(6,1)$}  (b)
        (d)  edge[bend left=20, dedge]  node[midway,below] {$(6,2)$}  (c)
        (d)  edge[bend right=20, dedge]  node[midway,below] {$(8,4)$}  (c);	
    \end{tikzpicture}
    
    \caption{Example of a temporal graph $\tg$. At each edge the availability and transition time is given as pair $(t, \lambda)$.}
    \label{fig:exampletga}
\end{figure} 

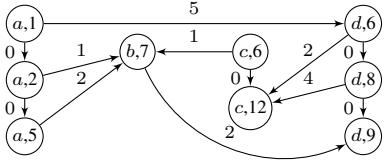
\begin{figure}[t]
    \centering
    \begin{tikzpicture}[scale=0.75]
        \node[vertex] (a5) at (0,0) {\scriptsize $a$,5};
        \node[vertex] (a2) at (0,1) {\scriptsize $a$,2};
        \node[vertex] (a1) at (0,2) {\scriptsize $a$,1};
        \node[vertex] (b7) at (2,1.5) {\scriptsize $b$,7};
        \node[vertex] (c6) at (4,1.5) {\scriptsize $c$,6};
        \node[vertex] (c12)at (4,0.5) {\scriptsize $c$,12};
        \node[vertex] (d6) at (6,2) {\scriptsize $d$,6};
        \node[vertex] (d8) at (6,1) {\scriptsize $d$,8};
        \node[vertex] (d9) at (6,0) {\scriptsize $d$,9};

        \path[->] 	
        (a1)  edge[dedge]  node[left] {\scriptsize$0$}   (a2)
        (a2)  edge[dedge]  node[left] {\scriptsize$0$}   (a5)
        (c6)  edge[dedge]  node[left] {\scriptsize$0$}   (c12)
        (d6)  edge[dedge]  node[left] {\scriptsize$0$}   (d8)
        (d8)  edge[dedge]  node[left] {\scriptsize$0$}   (d9)
        
        (a1)  edge[dedge]  node[above] {\scriptsize$5$}   (d6)
        (a2)  edge[dedge]  node[above] {\scriptsize$1$}   (b7)
        (a5)  edge[dedge]  node[above] {\scriptsize$2$}   (b7)
        (c6)  edge[dedge]  node[above] {\scriptsize$1$}   (b7)
        (b7)  edge[dedge,bend right=45]  node[above] {\scriptsize$2$}   (d9)
        (d6)  edge[dedge]  node[above] {\scriptsize$2$}   (c12)
        (d8)  edge[dedge]  node[above] {\scriptsize$4$}   (c12)
        ;	
    \end{tikzpicture}
    
    \caption{Example for the static time-respecting representation introduced in~\cite{gheibi2021effective} of the temporal graphs shown in \Cref{fig:exampletga}.}
    \label{fig:static_representation}
\end{figure}

\section{Design Goals, Architecture, and Temporal Graph Data Structures}
In the following, we will discuss the design goals and motivations for the high-level architecture of \tglib.
Furthermore, we introduce the data structures used for efficiently representing temporal graphs.

\subsection{General Architecture}
\Cref{fig:overview} shows a high-level view of the general architecture of \tglib. 
It consists of two main components: 1) the C++ template library and 2) the Python binding.
The C++ library provides the generic temporal graph data structures and contains the implementations of temporal graph algorithms. 
Furthermore, the C++ library offers functionality for IO, filtering, and data cleaning temporal graphs, which are common task necessary for real-world temporal graph data sets.

The Python interface allows non-C++-experts to use the efficient algorithms provided by \tglib.
We used Pybind11\footnote{\url{https://github.com/pybind/pybind11}} for generating the Python binding.

\subsection{Design Goals}
We developed \tglib under the following design goals:
\subsubsection{Performance and Efficiency}  
All running-time critical code is written in a modern C++ template library focusing on efficiency. 
Furthermore, many of the algorithms benefit from straightforward shared-memory parallelization; hence, when possible, we use OpenMP for loop parallelization and take advantage of modern shared-memory parallel processing capabilities. 
The lean temporal graph data structures are templates based and designed for running time and memory efficiency.

\subsubsection{Usability and Integration}
The C++ template library is provided as a platform-independent header-only library that can be easily integrated into existing or new C++ projects. 
Furthermore, we provide an easy-to-use Python interface with clear workflows for temporal graph algorithms and analysis to make \tglib widely available to researchers and practitioners.

\subsubsection{Exentability, Reuseability, and Sustainability}
The C++ source code of \tglib is based on a generic object-oriented modular designed to be easily extendable and reusable.
To this end, we use class templates for the temporal graph data structures to allow extension with customized data.
For example, weighted temporal graphs or temporal graphs with time-dependent node labels can be realized by providing 
corresponding edge or node datatypes.
Our implementation is fully documented, and we use unit tests and static code verification to ensure correctness and sustainability~\cite{roth2017clean}.

\begin{table}[t]
    \centering
    \caption{Overview of the temporal graph data structures.}
    \label{tab:datastructures}\renewcommand{\arraystretch}{1.0}
    \resizebox{1\linewidth}{!}{\setlength{\tabcolsep}{4pt}
        \begin{tabular}{lccc}\toprule
            \textbf{Name} & \textbf{Edge Type} & \textbf{Size} & \textbf{Reference}  \\ \midrule
            Temporal Edge Stream (\dsstr)         & Temporal & $\mathcal{O}(m)$   & \cite{wu2014path} \\
            Edge Incidence Lists (\dseil)         & Temporal & $\mathcal{O}(n+m)$ & \cite{oettershagen2020efficient} \\
            Time-Respecting Static Graph (\dstra) & Static   & $\mathcal{O}(n+m)$ & \cite{gheibi2021effective} \\
            Directed Line Graph (\dsdlg)          & Static   & $\mathcal{O}(m^2)$ & \cite{oettershagen2020temporal} \\
            Aggregated Graph (\dsagg)             & Static   & $\mathcal{O}(n^2)$ & \cite{holme2012temporal} \\
            \bottomrule
        \end{tabular}
    }
\end{table}

\begin{table}[t]
    \centering
    \caption{Overview of the implemented algorithms.}
    \label{tab:overview}\renewcommand{\arraystretch}{1.0}
    \resizebox{1\linewidth}{!}{\setlength{\tabcolsep}{5pt}
        \begin{tabular}{lllc}\toprule
            \textbf{Type} & \textbf{Algorithm} & \textbf{Data Structure} & \textbf{Reference} \\  
            \multirow{9}{1cm}{\emph{Distance}}
                                               & \multirow{3}{23mm}{Earliest Arrival/ Latest Departure}        
                                                    & \dsstr & \cite{wu2014path} \\
                                               &    & \dseil & \cite{dijkstra1959note,xuan2003computing }\\
                                               &    & \dstra &  New\\[1mm]
                                                
                                               & \multirow{3}{25mm}{Min.~Transition Sum/ Min.~Hops}   
                                                    & \dsstr &  \cite{wu2014path}\\
                                               &    & \dseil &  New\\
                                               &    & \dstra &  \cite{gheibi2021effective}\\[1mm]
                                               
                                               & \multirow{3}{2cm}{Fastest} 
                                                    & \dsstr & \cite{wu2014path}\\
                                               &    & \dseil & \cite{oettershagen2020efficient,oettershagen2022computing}\\
                                               &    & \dstra & \cite{gheibi2021effective}\\\midrule
            \multirow{11}{1cm}{\emph{Centrality}}  
                                               & \multirow{2}{25mm}{(In/Out-)Degree} 
                                                    & \dsstr & \cite{thompson2017static} \\
                                               &    & \dseil & \cite{thompson2017static} \\[1mm]
                                               & {Temporal Closeness} & \dsstr/\dseil/\dstra & \cite{santoro2011time}\\[1mm]

                                               & Top-k Closeness (Min. duration)         & \dseil & \cite{oettershagen2020efficient,oettershagen2022computing}\\[1mm]
                                               & Top-k Closeness (Shortest)         & \dseil & New \\[1mm]
                                               & Temporal Edge Betweeness & \dsdlg & New \\ [1mm]
                                               & Temporal Katz            & \dsstr & \cite{katztg}\\[1mm]
                                               & Temporal PageRank        & \dsstr & \cite{RozenshteinG16}\\[1mm]
                                               & Temporal Walk Centrality & \dsstr & \cite{temporalwalkcentrality}\\\midrule
               
            \multirow{7}{15mm}{\emph{Global/Local Properties}}  
                                               & Burstiness (Edges)       & \dsstr/\dseil & \cite{goh2008burstiness}\\[1mm]
                                               & Burstiness (Nodes)       & \dsstr/\dseil & \cite{goh2008burstiness}\\[1mm]
                                               & Temporal Clustering Coeff.   & \dsstr/\dseil & \cite{tang2009temporal}\\[1mm]
                                               & Temporal Diameter        & \dsstr/\dseil/\dstra &  \cite{holme2012temporal}\\[1mm]
                                               & Temporal Efficiency      & \dsstr/\dseil/\dstra &  \cite{tang2009temporal}\\[1mm]
                                               & Topological Overlap      & \dsstr/\dseil &   \cite{clauset2012persistence,tang2010small}\\  
            
            \bottomrule
        \end{tabular}
    }
\end{table}

\subsection{Temporal Graph Data Structures}
We implemented the temporal graph data structures listed in \Cref{tab:datastructures}.
The temporal graph data structures are provided as generic class templates.
\Cref{tab:datastructures} shows the worst-case sizes of the data structures using the implemented default classes.
In the following, we discuss the implemented data structures.
\subsubsection{Temporal Edge Streams (\dsstr)}\label{preliminaries:sec:algs:edgestreams}
The temporal graph is given as a sequence of its $m$ edges, chronologically ordered by the availability time of the edges in increasing order, with ties being broken arbitrarily~\cite{wu2014path}. 
This representation is often natural when events represented by the edges are sequentially recorded over time.
Algorithms for temporal edge streams usually pass over the edges in forward or backward sequential order and are often very efficient.
However, the \dsstr data structure can be disadvantageous for local computations, e.g., if we are interested in the immediate neighborhood of a single vertex, as we cannot directly access these neighbors.
\subsubsection{Edge Incidence Lists (\dseil)}\label{preliminaries:sec:edgeadjacencylist}
Here, the temporal graph consists of a set of temporal vertices, and each vertex has a list of temporal edges to its neighbors~\cite{oettershagen2020efficient}. 
The advantage of this representation is the local access to neighbors of a node, which is not directly possible in the temporal edge stream representation.

\subsubsection{Static Expansions}
Various static representations of temporal graphs offer different trade-offs between the size of the resulting static graph and the loss of temporal information. 
We implemented the following versions:

\paragraph{Time-Respecting Static Graph (\dstra)}
This representation was introduced in \cite{gheibi2021effective} and is an improved version of a data structure from~\cite{wu2014path}.
Here, the temporal graph is transformed into a static, i.e., non-temporal, graph.
The time-respecting static graph representation $S(\tg)=G_s=(V_s,E_s)$ of a temporal graph $\tg=(V,\tge)$ is defined as follows.
First, let $V_{o}(u)=\{(u,t)\mid (u,v,t,\lambda)\in\tge \}$, and $t_m(w) = \max\{t+\lambda \mid (v,w,t,\lambda) \}$ if $w$ has at least one incoming temporal edge.
We define $V'(u)=V_{o}(u)\cup \{(u,t_m(u))\}$ (or $V'(u)=V_{o}(u)$ if $u$ does not have an incoming edge) and $V_s=\bigcup_{u\in V}V'(u)$.
For each temporal edge $(u,v,t,\lambda)\in\tge$, we introduce a with $\lambda$ weighted static edge $((u,t), (v,t'))$ where $t'$ is the smallest arrival time at $v$ larger or equal to $t+\lambda$.
Furthermore, for each $u\in V$, the vertices in $V'(u)$ are connected with zero weighted edges in ascending order.
\Cref{fig:static_representation} shows the \textsc{Trs} of the temporal graph shown in \Cref{fig:exampletga}.

\paragraph{Directed Line Graph (\dsdlg)}
The directed line graph expansion has been previously used for survivability and reliability analysis~\cite{khanna2020two,liang2016survivability}. 
In~\cite{oettershagen2020classifying,oettershagen2020temporal}, the authors used the \dsdlg for lifting static graph kernels to the temporal domain.
In a recent work~\cite{temporalwalkcentrality}, the authors use the \dsdlg for algebraic weighted walk counting.
Given a temporal graph $\tg=(V,\tge)$, the \emph{directed line graph}
$\mathit{DL}(\tg)=(V',E')$ is the directed graph, where  
every temporal edge $(u,v,t,\lambda)$ in $\tge$ is represented by a vertex $n^{t}_{uv}$, 
and there is an edge from $n^{t}_{uv}$ to  $n^{s}_{xy}$ if $v=x$ and $t+\lambda\leq s$.
%

\paragraph{Aggregated Static Graph (\dsagg)}
Given a temporal graph $\tg$, removing all time stamps and traversal times, and merging resulting multi-edges, we obtain the \emph{aggregated}, or \emph{underlying static}, graph $A(\tg)=(V,E_s)$ with $E_s=\{(u,v)\mid(u,v,t,\lambda)\in\tge\}$. 
The edges can be weighted depending on the number of temporal edges, e.g., using the contact frequency, i.e., $\phi((u,v))=|\{(u,v,t,\lambda)\in\tge\}|$.
The aggregated graph can be much smaller than the temporal graph as its number of edges is in $\mathcal{O}(n^2)$. 
However, it does not preserve the temporal information of the network.

\section{Implemented Algorithms}
\Cref{tab:overview} gives an overview of the implemented algorithms and the underlying data structures.
All implemented algorithms can be restricted to only consider a given time interval $I$ without increasing the running times. 
In the following, we discuss the implemented algorithms.

\subsection{Temporal Paths, Reachability, and Distances}
Finding temporal paths, deciding reachability, and determining temporal distances are essential tasks in various applications and scenarios, e.g., in the computation of temporal centrality measures~\cite{braha2009time,oettershagen2020efficient}, solving time-dependent transportation problems~\cite{gendreau2015time,idri2017new,pyrga2008efficient}, or in the simulation and analysis of epidemics~\cite{enright2018epidemics,lentz2016disease}.

\begin{definition}
    A \emph{temporal walk} in a temporal graph $\tg$ is an alternating sequence $(v_1, e_1,\ldots,e_k,v_{k+1})$ of vertices and temporal edges connecting consecutive vertices 
    where for $1\leq i < k$, $e_i=(v_i,v_{i+1},t_i,\lambda_i)\in \tge$, and $e_{i+1}=(v_{i+1},v_{i+2},t_{i+i},\lambda_{i+1})\in \tge$ the time $t_i+\lambda_{i}\leq t_{i+1}$ holds. 
    A \emph{temporal path} $P$ is a temporal walk in which each vertex is visited at most once. 
\end{definition}
For notational convenience, we omit vertices. 
The length of a temporal walk $\omega$ is the number of edges it contains, and we denote it with $|\omega|$. 
Let $\omega=(e_1,\ldots,e_\ell)$ be a temporal walk in a temporal graph $\tg$.
The \emph{starting time} of $\omega$ is $s(\omega)=t_1$, the \emph{arrival time} is $a(\omega)=t_\ell+\lambda_\ell$, and
the \emph{duration} is $d(\omega)=a(\omega)-s(\omega)$. 
Finally, we define $l(\omega)=\sum_{i=1}^{\ell}\lambda_i$.

For example, in \Cref{fig:exampletga}, there are three paths between vertices $a$ and $d$. 
The first one consists of only the edge $P_1=((a,d,1,5))$ and with $d(P_1)=5$. 
The second is $P_2=((a,b,2,1),(b,d,7,2))$ with $d(P_2)=7$. 
And, path three $P_3=((a,b,5,2),(b,d,7,2))$ with $d(P_3)=4$. 

There are several optimality criteria for temporal paths used in the literature, and 
we distinguish the following.
\begin{definition}\label{preliminaries:def:opt}
    Let $\tg$ be a temporal graph and $\mathcal{P}$ be the set of all temporal paths in $\tg$.
    A $(s,z)$-path\footnote{We use $z$ instead of $t$ as the target vertex because we use $t$ to denote a time stamp.} $P\in\mathcal{P}$ is 
    \begin{itemize}
        \item an \emph{earliest arrival} path if there is no other $(s,z)$-path $P'\in\mathcal{P}$ with $a(P')<a(P)$,
        \item a \emph{latest departure} path if there is no other $(s,z)$-path $P'\in\mathcal{P}$ with $s(P')>s(P)$,
        \item a \emph{minimum duration}, or \emph{fastest}, path if there is no other $(s,z)$-path $P'\in\mathcal{P}$ with $d(P')<d(P)$,  
        \item a \emph{shortest} path if there is no other $(s,z)$-path $P'\in\mathcal{P}$ with $l(P')<l(P)$, and
        \item a \emph{minimum hops} path if there is no other $(s,z)$-path $P'\in\mathcal{P}$ with $|P'|<|P|$.
    \end{itemize}
\end{definition}

For the example in \Cref{fig:exampletga}, $P_3$ is the only fastest $(a,d)$-path. Notice that the subpath $P'_3=((a,b,5,2))$ is not a fastest $(a,b)$-path. The only fastest $(a,b)$-path consists of edge $(a,b,2,1)$ and has a duration of one.

We provide algorithms for determining the temporal distances of \Cref{preliminaries:def:opt} for the different data structures, see~\Cref{tab:overview}.
The reason for providing the algorithms for different data structures is that depending on the topology of the graph, 
different algorithms can be more efficient than others, see, e.g.,~\cite{wu2014path,oettershagen2020efficient,gheibi2021effective}.
We additionally introduce a new shortest paths algorithm for the \dseil data structure and a new earliest arrival algorithm for the 
\dstra data structure\footnote{A formal description of the algorithms with correctness and complexity proofs will be in an extended version of this paper.}. 
For each temporal distance, we provide algorithms to obtain an optimal temporal path and the temporal diameter, which is 
defined as the maximum optimal temporal distance between any two (reachable) vertices in the network~\cite{calamai2021computing}.

\subsection{Centrality Measures}
The centrality of a node (edge) in a network quantifies its structural importance. Various functions can be used to measure node (edge) centrality by assigning values corresponding to some measurement of importance to each node (edge), where the informative value must be assessed based on a research question. 
For introductions of centrality approaches, see, e.g.,~\cite{das2018study,landherr2010critical,rodrigues2019network,Saxena2020}. 
\tglib provides the following centrality measures designed explicitly for temporal graphs.


\subsubsection{Temporal Closeness}
Due to the differences in reachability and optimality in temporal graphs, several versions of temporal closeness have been suggested, see, e.g.,~\cite{pan2011path,crescenziMM20topk,oettershagen2020efficient}.
Using the optimal distance computations for earliest arrival paths, fastest paths, etc., we provide four different versions of harmonic temporal closeness
defined as
\[
C(u)=\sum_{v\neq u\in V}\frac{1}{d(u,v)},
\]
where we define $1/\infty=0$ for non-reachable vertices.
Furthermore, we provide the top-$k$ approach introduced in~\cite{oettershagen2020efficient} for finding the $k$ highest closeness centrality values and the corresponding vertices. The authors of~\cite{oettershagen2020efficient} only introduced their top-$k$ algorithm for harmonic closeness wrt.~to the minimum duration distance. 
We provide a new additional implementation for the minimum transition times distance.



\subsubsection{Temporal Edge Betweenness}
Similar to the static edge betweenness~\cite{brandes2008variants}, the temporal edge betweenness is an edge centrality measure and quantifies the importance of the temporal edges in terms of the shortest temporal paths crossing the temporal edge.
It can be computed by counting the shortest paths in the directed line graph representation due to the one-to-one mapping of walks in $\tg$ and $DL(\tg)$~\cite{oettershagen2020temporal}.

\subsubsection{Temporal Katz Centrality}
The \emph{Katz centrality} introduced in \cite{katz1953new} measures vertex importance in terms of the number of random walks starting (or arriving) at a vertex, down-weighted by their length. 
The authors of~\cite{katztg,grindrod2011communicability} adapt the walk-based Katz centrality to temporal graphs.

\subsubsection{Temporal PageRank Centrality}
Rozenshtein and Gionis~\cite{RozenshteinG16} incorporate the temporal character in the definition of the static PageRank originally introduced by~\cite{Page1999}. They obtain a temporal PageRank by replacing walks with temporal walks. 

\subsubsection{Temporal Walk Centrality}
Temporal Walk Centrality is a recently proposed centrality measure that aims to rank the vertices according to their ability to obtain and distribute information~\cite{temporalwalkcentrality}.

\subsection{Further Local and Global Properties}
Moreover, we implemented the following local and global temporal graph properties.

\subsubsection{Burstiness}
Burstiness measures how much a sequence of contacts $\tau$ (of a single node or between a pair of nodes) deviates from the memoryless random Poisson process~\cite{holme2012temporal}.
It is defined as
\[
B(\tau)=\frac{\sigma_\tau - m_\tau}{\sigma_\tau + m_\tau} \in [-1,1],
\]
where $\sigma_\tau$ and $m_\tau$ denote the standard deviation and mean of the inter-contact times $\tau$, respectively~\cite{goh2008burstiness}.
A value close to one indicates a very \emph{bursty} sequence, and a value close to minus one a more periodic sequence.

\subsubsection{Temporal Clustering Coefficient}
The temporal clustering coefficient is defined as 
\[
C_C(u) = \frac{\sum_{t\in T(\tg)} \pi_t(u)}{|T(\tg)|{|N(u)| \choose 2}},
\]
where $\pi_t(u)=|\{(v,w,t,\lambda)\in\tge\mid v,w\in N(u)\}|$~\cite{tang2009temporal}.  

\subsubsection{Temporal Efficiency}
The temporal efficiency is a global statistic based on the temporal closeness values of the nodes~\cite{tang2009temporal}. 
It is defined as 
\[
T_{\text{eff}}(\tg)=\frac{1}{n(n-1)}\sum_{u\in V(\tg)}\sum_{v\neq u\in V(\tg)}\frac{1}{d(u,v)}
\]
with $d(u,v)$ being a temporal distance and $1/\infty=0$ in case of non-reachable vertices.

\subsubsection{Topological Overlap}

The topological overlap of a node is defined as
\[
T_{\text{to}}(u)=\frac{1}{T(\tg)}\sum_{t=1}^{T(\tg)}\frac{\sum_{v\in{N}(u)}\phi^t_{uv}\phi^{t+1}_{uv}}{\sqrt{\sum_{v\in{N}(u)}\phi^t_{uv}\sum_{v\in{N}(u)}\phi^{t+1}_{uv}}},
\]
where $\phi^{t}_{uv}=1$ iff.~there exists a temporal edges between $u$ and $v$ at time $t$ and zero otherwise~\cite{clauset2012persistence,tang2010small}.
In case that the denominator equals one, we define $T_{\text{to}}(u)=1$.
And the global topological overlap is defined as
$
T_{\text{to}}(\tg)=\frac{1}{n}\sum_{u\in V}T_{\text{to}}(u)
$.
The topological overlap lies in the range between zero and one. 
A value close to zero means many edges change between consecutive time steps, and a value close to one means there are often only a few changes.



\section{Comparison to Related Software}
There are several popular graph libraries designed for conventional static graphs, e.g., Networkit~\cite{staudt2016networkit}, OGDF~\cite{chimani2013open}, LEMON~\cite{dezsHo2011lemon}, or Boost graph~\cite{siek2002boost}. 
However, they are not designed to handle the peculiarities of temporal graphs, e.g., they do not support algorithms that respect the temporal restrictions in temporal walks and paths.
The SNAP library provides various algorithms for temporal graphs, like counting specific temporal motifs~\cite{paranjape2017motifs}.
We are not aware of a dedicated C++ library for temporal graphs. For Python, the Teneto library is a dedicated temporal graph library supporting various analytical methods~\cite{thompson2017static}.
However, the library focuses on analyzing small networks obtained from fMRI brain scans and does not support transition times on the edges. Moreover, it does not perform well on mid-size to large temporal graphs as its mainly based on Python code and matrix-based computations.
Finally, Teneto is published under the, compared to the MIT license, more restrictive GNU GPL3 license.

\section{Python Interface}
\label{sec:python-usage}
TGLib is available as a precompiled Python package on the Python Package Index (PyPI).\footnote{\url{https://pypi.org/project/temporalgraphlib/}} This enables straightforward installation and usage within Python environments.
%
 The library can be installed via \texttt{pip} as follows:
\begin{lstlisting}[language=bash,style=mystyle,label=example]
	pip install temporalgraphlib
\end{lstlisting}

\noindent After installation, the functionality of TGLib is accessible through the \texttt{pytglib} module. For example:
\begin{lstlisting}[language=Python,style=mystyle,label=example,caption={Example Python Usage}]
	import pytglib as tgl
	
	# Load a temporal graph from an ordered edge list
	tgs = tgl.load_ordered_edge_list("example.tg")
	
	# Compute and print basic statistics
	stats = tgl.get_statistics(tgs)
	print(stats)
\end{lstlisting}

\noindent This interface provides efficient access to the core temporal graph algorithms and is particularly useful for rapid prototyping and exploratory analysis.

\section{Example Use-Case}
For this example use-case, we load two real-world data sets and compare different variants of temporal closeness.
The first one is the \emph{AskUbuntu}, a network consisting of interactions on the stack exchange website \emph{Ask Ubuntu}~\cite{paranjape2017motifs}.
The second data set is the \emph{Enron} email network between employees of a company~\cite{klimt2004enron}.
To obtain the basic statistics, we load the temporal graph and call the \texttt{get\_statistics} function (see~\Cref{example}). \Cref{tab:stats} shows (a subset) of the returned statistics. Note that Teneto is unable to load these data sets due to its temporal graph representation as a sequence of adjacency matrices and the resulting out-of-memory error (6.13 EiB for \emph{Enron}).
Next, we compute the closeness centrality with respect to minimum duration distance and earliest arrival time.
We compare the obtained rankings using Kendall's $\tau$ rank correlation using SciPy\footnote{\url{https://scipy.org/}}.
We obtain correlations of $0.79$ for \emph{AskUbuntu} and $0.94$ for \emph{Enron}, showing that the two types of temporal closeness are strongly 
correlated in both graphs.
\begin{lstlisting}[language=Python,style=mystyle,label=example,caption={Example Use-case}]
    import pytglib as tgl    
    import scipy.stats as ss # for correlation 
    tgs = tgl.load_ordered_edge_list("datasetname")
    stats = tgl.get_statistics(tgs)
    print(stats)
    closeness_fastest = tgl.temporal_closeness(tgs, tgl.Distance_Type.Fastest) 
    closeness_ea = tgl.temporal_closeness(tgs, tgl.Distance_Type.Earliest_Arrival)
    tau, p_value = ss.kendalltau(closeness_fastest, closeness_ea)
\end{lstlisting}
\begin{table}[t]
    \centering
    \caption{Overview of the temporal graph statistics.}
    \label{tab:stats}\renewcommand{\arraystretch}{1.0}
    \resizebox{1\linewidth}{!}{\setlength{\tabcolsep}{7pt}
        \begin{tabular}{lcccccc}\toprule
            \textbf{Data set} & $|V|$ & $|\tge|$ & $|T(\tg)|$ & $|E_s|$ & $\delta^-_{max}$ & $\delta^+_{max}$ \\ \midrule
            \emph{AskUbuntu} & 159\,316& 964\,437& 960\,866&596\,933 &4\,926 & 8\,729\\
            \emph{Enron}     & 87\,101 & 1\,147\,126 & 220\,312 & 321\,288 & 6\,165 & 32\,613 \\
            \bottomrule
        \end{tabular}
    }
\end{table}



\section{Open-Source Development and License}
\tglib free software licensed under the permissive MIT License.
It is available at \textcolor{blue}{\url{https://gitlab.com/tgpublic/tglib}}.
We aim to encourage a diverse community, including network researchers, data mining practitioners, and algorithm engineers, to use \tglib and contribute to the open-source development. 

\section{Conclusion and Future Work}

We introduced the open-source toolkit \tglib, a C++ library for efficient temporal graph analysis featuring an easy-to-use and accessible Python front-end.
So far, we have implemented a wide range of algorithms for distance, centrality, and analytical computations based on various efficient temporal graph data structures. \tglib offers researchers, practitioners, and students convenient access to temporal graph algorithms. Furthermore, it offers a unified and accessible approach for reproducibility and comparability. 
We are actively working to integrate further and future methods and algorithms into our library. 
Using the permissive MIT license, we hope that \tglib will be used and extended by the temporal graph community.


\end{document}